\documentclass[11pt]{article}
\usepackage{mymoriond,epsfig,url}

\def\Journal#1#2#3#4{{#1} {\bf #2}, #3 (#4)}

\def\EPJCd{{\em Eur.\ Phys.\ J.\ } direct C}
\def\EPJC{{\em Eur.\ Phys.\ J.\ } C}
\def\JPG{\em J.\ Phys.\ G: Nucl.\ Part.\ Phys.\ }
\def\JHEP{\em J.\ High Energy Phys.\ }

\def\NPB{{\em Nucl.\ Phys.\ } B}
\def\PLB{{\em Phys.\ Lett.\ }  B}
\def\PRL{\em Phys.\ Rev.\ Lett.\ }
\def\PRD{{\em Phys.\ Rev.\ } D}
\def\RMP{\em Rev.\ Mod.\ Phys.\ }

\def\be{\begin{equation}}
\def\ee{\end{equation}}
\def\bea{\begin{eqnarray}}
\def\eea{\end{eqnarray}}

\newcommand{\babar}{\mbox{\slshape B\kern-0.1em{\small A}\kern-0.1em
    B\kern-0.1em{\small A\kern-0.2em R}}}

\begin{document}
\vspace*{2cm}
\title{REVIEW OF EXPERIMENTAL RESULTS ON RARE RADIATIVE, 
 SEMILEPTONIC AND LEPTONIC B DECAYS}

\author{ Thomas SCHIETINGER }

\address{Laboratory for High-Energy Physics,
Ecole Polytechnique F\'ed\'erale,\\ CH-1015 Lausanne, Switzerland}

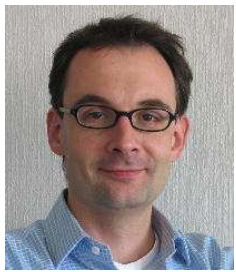
\maketitle\abstracts{
We review recent experimental progress in the domain of rare radiative, 
semileptonic and leptonic $B$ decays.
The statistical precision attained for these decays has reached
a level where they start to impose meaningful constraints on 
the Cabibbo-Kobayashi-Maskawa matrix, which are complementary to
those obtained from hadronic decays.
While the current data indicate no deviations from Standard Model
predictions, there is still some room for new physics in these
decays.}

Rare $B$ decays to photons and leptons are among the cleanest probes
of the flavour sector of the Standard Model (SM) available to present
experiments.
In particular, these decays have the potential of revealing the existence
of new couplings not present in the SM.
The field has seen tremendous progress in the last few years, 
thanks to the large data samples accumulated by the two 
asymmetric $B$ factories, \babar\ and Belle operating at the PEP-II and KEKB $e^+e^-$ colliders, 
respectively, at a centre-of-mass energy corresponding to the mass of the 
$\Upsilon(4S)$ resonance.

Please note that {\em all branching fractions given in this review are
in units of 10$^\mathit{-6}$} and all limits are to be understood at 
90\% confidence level.
The symbol $\ell$ stands for all three charged leptons ($e$, $\mu$ and $\tau$),
unless otherwise specified.
The first and second uncertainties quoted on measurements are statistical 
and systematic, respectively.

\section{Radiative penguin decays}
\label{sec:radpeng}

\subsection{$b\rightarrow s\gamma$ inclusive}\label{subsec:b2sgincl}

The primary motivation for inclusive $b\rightarrow s\gamma$ measurements
is the search for effects from physics beyond the SM.
The branching fraction directly probes the Wilson coefficient
$C_7$ (see also Sec.~\ref{sec:slpeng}), whereas the direct CP asymmetry 
is sensitive to new phases appearing in the decay loop.
Moreover, $b\rightarrow s\gamma$ is an ideal laboratory for studying
the dynamics of the $b$-quark inside the $B$ meson:
since the motion of the $b$-quark inside the $B$ meson is universal,
information gained from a measurement of the energy spectrum of the
emitted photon in $b\rightarrow s\gamma$ is applicable to other
processes, for instance semileptonic decays.

Experimentally, two methods are used to extract the $b\rightarrow
s\gamma$ (more precisely: $B\rightarrow X_s\gamma$) signal.
In the first, fully inclusive method, events containing a hard 
photon consistent with $B\rightarrow X_s\gamma$ are selected.
The resulting very large backgrounds, primarily from $q\bar{q}$ 
continuum events are suppressed as much as possible using sophisticated
techniques based on event-shape and energy-flow variables, 
and then subtracted by use of off-resonance data
taken below the $\Upsilon(4S)$ resonance. 
Particularly effective suppression of continuum backgrounds
is afforded by the requirement of a high-$p_t$ lepton, signaling
the semileptonic decay of the accompanying $B$ meson, as applied by 
\babar.
In the second, semi-inclusive method, the $B\rightarrow X_s\gamma$
rate is determined from a sum of exclusive modes with an extrapolation
procedure to take account of the unobserved modes (modes containing
$K_L$ for instance). This extrapolation, which is based on the assumption of
isospin symmetry and Monte Carlo (MC) simulation, contributes the largest
systematic uncertainty in this method.

The most recent measurements are summarized in Table~\ref{tab:b2sgincl}.
The Heavy Flavor Averaging Group~\cite{HFAG} (HFAG) has recently issued a new
average using a common shape function~\cite{PRD73-073008-2006} for the extrapolation to lower
photon energies and taking into account the correlated error from 
$b\rightarrow d\gamma$:
\be
\mathcal{B}(b\rightarrow s\gamma) = 355 \pm 24^{+9}_{-10}\pm 3 ,
\ee
for a photon cut-off energy $E_\gamma >$ 1.6 GeV. 
The errors on this average are experimental (combined statistical and systematic),
systematic due to the shape function, and systematic due to the $d\gamma$ fraction.
Comparing with the corresponding value from theory,\cite{NPB631-219-2002} 
$\mathcal{B}(b\rightarrow s\gamma) = 357 \pm 30$, we cannot help being
impressed by the agreement---or depressed, if the goal is to find new physics!

\begin{table}[t] 
\caption{Recent $b\rightarrow s\gamma$ inclusive branching fraction measurements, as
reported by the experiments.
The errors are statistical, systematic, and shape-function systematic
(from the extrapolation below the photon energy cut-off). }
\begin{center}
\begin{tabular*}{\textwidth}{@{\extracolsep{\fill}}lllll}
\hline
Collaboration                 & method     & $\mathcal{B}$ measurement                
   & \hspace{-8mm}$E_\gamma$ cut-off [GeV] & comment \\
\hline
\vspace{-0.4cm}&&&&\\
CLEO~\cite{PRL87-251807-2001}  & incl.      & $321\pm43\pm27^{+18}_{-10}$              & 2.0  & \\
Belle~\cite{PRL93-061803-2004} & incl.      & $355\pm32^{+30}_{-31}\mbox{}^{+11}_{-7}$ & 1.8  & \\
Belle~\cite{PLB511-151-2001}   & semi-incl. & $336\pm53\pm42^{+50}_{-54}$              & 2.24 & 16 modes \\
\babar~\cite{hep-ex/0507001}   & incl.      & $367\pm29\pm34\pm29^{a}$                 & 1.9 & lepton-tagged \\
\babar~\cite{PRD72-052004-2005}& semi-incl. & $335\pm19^{+56}_{-41}\mbox{}^{+4}_{-9}$  & 1.9 & 38 modes\\
\vspace{-0.4cm}&&&&\\
\hline
\label{tab:b2sgincl}
\end{tabular*}
\end{center} \vspace{-5mm}
{\footnotesize $^{a}$ The branching fraction is not extrapolated below the photon energy cut-off.
The third error in this case refers \vspace{-0.8mm} \\ 
to a model dependence in the efficiency evaluation.\vspace{-4mm}}
\end{table}

Since significant improvements in precision cannot be expected on 
either the experimental or the theoretical side, the focus in 
$b\rightarrow s\gamma$ studies has shifted towards the measurement
of the photon energy spectrum, which gives direct experimental 
access to parameters that can be related to the mass and momentum
of the $b$-quark inside the $B$-meson and are therefore of great 
interest in many areas of $B$ physics and beyond.
CLEO,\cite{PRL87-251807-2001}
\babar~\cite{hep-ex/0507001} and 
Belle~\cite{PRL93-061803-2004} 
have published spectra in the $\Upsilon(4S)$
rest frame that are, in that order, based on increasing 
data samples and going to lower and lower photon energy cutoffs.
The semi-inclusive analysis published by \babar~\cite{PRD72-052004-2005} 
allows a spectrum measurement in the $B$ rest frame (via the invariant 
mass of the strange hadronic recoil system). 
The spectrum obtained by this method is therefore free from the 
smearing due to the $B$ momentum and profits from a much better
energy resolution provided by the tracks of the hadronic recoil
rather than the electromagnetic calorimeter.

As for the CP asymmetry measured in inclusive $b\rightarrow s\gamma$,
neither \babar~\cite{PRL93-021804-2004} 
($A_{CP} = (2.5\pm5.0\pm1.5)\%$)
nor Belle~\cite{PRL93-031803-2004} 
($A_{CP} = (0.2\pm5.0\pm3.0)\%$) 
reports significant asymmetries from their semi-inclusive analyses,
in accordance with SM predictions.
A new result from \babar\ based on its fully inclusive lepton-tagged
analysis~\cite{hep-ex/0507001} (which does not distinguish $b\rightarrow s\gamma$ from 
$b\rightarrow d\gamma$) gives $(-11.0\pm11.5\pm1.7)\%$ for the 
CP-asymmetry in $b\rightarrow [s+d]\gamma$.
Note that in the limit of U-spin symmetry, this asymmetry is strictly 
zero by unitarity in the SM.\cite{NPB-704-56-2005}

\subsection{$b\rightarrow s\gamma$ exclusive}\label{subsec:b2sgexcl}

On the exclusive front, the kaon resonance modes 
$B\rightarrow K^*(892)\gamma$ 
(CLEO,\cite{PRL84-5283-2000}
\babar,\cite{PRD70-112006-2004} and
Belle~\cite{PRD69-112001-2004}),
$B\rightarrow K_1(1270)\gamma$, (Belle~\cite{PRL94-111802-2005}) and
$B\rightarrow K_2^*(1430)\gamma$ 
(CLEO,\cite{PRL84-5283-2000} 
Belle,\cite{PRL89-231801-2002} and
\babar~\cite{PRD70-091105-2004})
are by now well established.
The measured branching fractions are in good agreement with, yet
more precise than, theoretical predictions.

The list of established decays of the type 
$B\rightarrow K^{(*)}X\gamma$, where $X$ stands
for one or more flavourless mesons is also growing longer:
Apart from  
$K\pi\gamma$,
$K\pi\pi\gamma$ and 
$K^*\pi\gamma$,\cite{PRL89-231801-2002}
it also includes
$K\eta\gamma$ (Belle~\cite{PLB610-23-2005} and \babar~\cite{hep-ex/0603054}) and
$K\phi\gamma$ (Belle~\cite{PRL92-051801-2003}), 
where most of these channels have been found to be produced
via resonances.
Most noteworthy among the newer results are the \babar\ analyses
on $B\rightarrow K\eta\mbox{}^{(}\mbox{}'\mbox{}^{)}\gamma$ 
and $B\rightarrow K\pi\pi\gamma$, both
based on 232M $B\overline{B}$ pairs:
 \babar\ reports the first observation~\cite{hep-ex/0603054} 
of the neutral decay $B^0\rightarrow K^0\eta\gamma$ and gives 
the very first limits on the channel $B\rightarrow K\eta'\gamma$, 
which is expected to be suppressed with respect to $B\rightarrow 
K\eta\gamma$ due to the well-known destructive interference of 
penguin diagrams. 
The $B\rightarrow K\pi\pi\gamma$ study~\cite{hep-ex/0507031} 
performed by \babar\ yielded the first observations of the channels
$B^0\rightarrow K^+\pi^-\pi^0\gamma$ and 
$B^+\rightarrow K^0\pi^+\pi^0\gamma$.
These channels are of interest for a measurement of the polarization
of the photon emitted in the $b\rightarrow s\gamma$ process, see
Sec.~\ref{subsec:b2sgpol}. 

The search for direct CP violation in exclusive radiative decays
has reached the few-percent level in the channel 
$B\rightarrow K^*(892)\gamma$.\cite{PRD70-112006-2004,PRD69-112001-2004}
The latest HFAG average~\cite{HFAG} reads 
$A_{CP}(B\rightarrow K^*(892)\gamma)$ = $(-1.0\pm2.8)\%$.
A new result from \babar\ is
$A_{CP}(B^+\rightarrow K^+\eta\gamma) = 
(-9\pm12\pm1)\%$.\cite{hep-ex/0603054} 

In short, Belle has been leading in this domain, but \babar\ is
rapidly catching up!

\subsection{Photon polarization in $b\rightarrow s\gamma$}
\label{subsec:b2sgpol}

The polarization of the photon emitted in the $b\rightarrow s\gamma$
transition provides an important test of the SM, which predicts
a mostly left-handed photon.\cite{PRL79-185-1997,PRD71-011504-2005}
The two most promising methods to access this polarization experimentally
at the $B$ factories rely either on $B^0$-$\overline{B}\mbox{}^0$ 
interference~\cite{PRL79-185-1997} or on interference effects in decays 
to higher kaon resonances~\cite{PRL88-051802-2002} producing
$K\pi\pi^0$. 
In the first method, the time-dependent interference between 
$B^0$ and $\overline{B}\mbox{}^0$ decaying to the final state 
$K_S\pi^0\gamma$ (where $K_S\pi^0$ may or may not be 
resonant~\cite{PRD71-076003-2005}) is expected to be suppressed for
a polarized photon, since in that case the final state is no longer
CP invariant.
The reconstruction of the $K_S\pi^0\gamma$ vertex is experimentally
challenging, but possible by intersecting the reconstructed $K_S$
momentum direction with the beam envelope.
The most recent measurements of the $S$ parameter describing the 
interference (equivalent to $\sin(2\phi_1)$ or $\sin(2\beta)$ in
$b\rightarrow c\bar{c}s$ transitions) indeed seem to favour a small
value (\babar~\cite{PRD72-051103-2005} measures 
$S_{K^{*0}\gamma} = -0.21\pm0.40\pm0.05$ on a 232M $B\overline{B}$ 
sample, Belle~\cite{hep-ex/0507059} has
$S_{K_S\pi^0\gamma} = 0.08\pm0.41\pm0.10$ 
from 386M $B\overline{B}$), 
but are not precise enough yet to put significant constraints on the 
photon polarization.
As for the other method, the recent \babar\ study~\cite{hep-ex/0507031} 
of $B\rightarrow K\pi\pi^0\gamma$ reveals a resonance structure that
is rather difficult to disentangle, thus precluding a measurement
of the photon polarization at this time.
In particular there is no clear evidence yet of the $K_1(1400)$ resonance, 
a prerequisite for the method.

We conclude that constraining the photon polarization in the
$b\rightarrow s\gamma$ transition remains an elusive goal for
the $B$ factories even with roughly half their expected final
statistics available.
Since other proposed methods, based on photon conversion~\cite{JHEP0006-029-2000}
or interference with radiative charmonium decays~\cite{PLB634-403-2006} require
even larger data samples, we may well have to wait for a Super-$B$ factory
to obtain a definitive answer.
On the other hand, this may represent a good opportunity for 
experiments at the LHC, where radiative decays of 
$\Lambda_b$ baryons~\cite{JPGNPP24-979-1998} present an interesting 
approach, in particular if they are sufficiently 
polarized.\cite{PRD65-074038-2002}
Further constraints on right-handed currents in $b\rightarrow s\gamma$ may
come from $B\rightarrow K^*\ell^+\ell^-$ decays.\cite{PLB442-381-1998}
First efforts to this end have been presented by \babar\ very 
recently.\cite{PRD73-092001-2006}

\subsection{$b\rightarrow d\gamma$}\label{subsec:b2dgexcl}

The quark transition $b\rightarrow d\gamma$ is the 
Cabibbo-Kobayashi-Maskawa-(CKM-)suppressed counterpart of $b\rightarrow s\gamma$, 
therefore expected at a rate smaller by a factor $|V_{td}/V_{ts}|^2 
\approx 0.04$.
An additional weak annihilation diagram also contributes to the exclusive
decay $B^+\rightarrow \rho^+\gamma$, slightly complicating the 
extraction of $|V_{td}/V_{ts}|^2$ from branching fraction measurements.

In 2005, Belle announced the first observation~\cite{hep-ex/0506079} of the 
$b\rightarrow d\gamma$ transition with a data sample equivalent to 
386M $B\overline{B}$ pairs.
The analysis makes use of very sophisticated background-suppression techniques
including $\pi^0$ and $\eta$ rejection, the requirement of spatial vertex
separation between the two $B$ mesons and application of a flavour-tagging
algorithm to distinguish $B\overline{B}$ from continuum events.
The background discriminating variables are combined with an event-shape
Fisher discriminant to form a signal-background likelihood ratio.
A flavour-tag-quality dependent cut is then applied on this likelihood ratio.
The mode $B^0\rightarrow\rho^0\gamma$ is observed with a significance of 
$5.2\sigma$, $\mathcal{B}(B^0\rightarrow\rho^0\gamma) = 
1.25^{+0.37}_{-0.33}\mbox{}^{+0.07}_{-0.06}$, whereas the corresponding
charged decay is only seen at $1.6\sigma$ significance,
$\mathcal{B}(B^+\rightarrow\rho^+\gamma) = 0.55^{+0.42}_{-0.36}\mbox{}^{+0.09}_{-0.08}$,
in apparent contradiction to the expectation from isospin invariance.
The probability of observing an isospin violation this large or larger
is evaluated to be 4.9\%.
Theoretically, an isospin violation of $\pm10\%$ is expected. 
A combined fit of $\rho^+\gamma$, $\rho^0\gamma$ and $\omega\gamma$
candidate events for an isospin-averaged rate gives 
$\overline{\mathcal{B}}(B\rightarrow [\rho,\omega]\gamma) = 
1.32^{+0.34}_{-0.31}\mbox{}^{+0.10}_{-0.09}$, with a significance
of $5.1\sigma$.
From this and by use of the formula \cite{PLB595-323-2004}
\be
\frac{\overline{\mathcal{B}}(B\rightarrow [\rho,\omega]\gamma)}
     {\mathcal{B}(B\rightarrow K^*\gamma} =
\left|\frac{V_{td}}{V_{ts}}\right|^2
\left(\frac{1-m_{\rho,\omega}^2/M^2_B}{1-m_{K^*}^2/M^2_B}\right)^3
\zeta^2 (1+\Delta R),
\ee
where $\zeta = 0.85\pm0.10$ is the relevant form factor ratio and
$\Delta R = 0.1\pm0.1$ parameterizes the $SU(3)$ breaking due to the 
weak annihilation diagram,\footnote{More advanced calculations of  
$\zeta$ and $\Delta R$ have recently become available.\cite{JHEP-0604-046-2006}}
Belle obtains 
$|V_{td}/V_{ts}| = 0.199^{+0.026}_{-0.025}\mathrm{(exp.)}$
$^{+0.018}_{-0.015}\mathrm{(theor.)}$,
in good agreement with global CKM fits.

\babar\ has performed a similar analysis on a data sample containing
211M $B\overline{B}$ pairs.\cite{PRL94-011801-2005}
They use a neural network to combine various background suppression
variables.
The obtained branching fractions are not significant. 
For the isospin-averaged branching fraction, the upper limit
$\overline{\mathcal{B}}(B\rightarrow [\rho,\omega]\gamma) < 1.2$
is given, which translates to $|V_{td}/V_{ts}| < 0.19$.

\section{Semileptonic penguin decays}
\label{sec:slpeng}

The physics of~\footnote{Throughout this section,
$\ell$ stands for $\mu$ and $e$ only.}
 $b\rightarrow s\ell^+\ell^-$ is governed by the Wilson
coefficients $C_7$, $C_9$ and $C_{10}$, which describe the strengths of
the corresponding short-distance operators in the effective Hamiltonian,
i.e.\ the electromagnetic operator $O_7$ and the semileptonic vector 
and axialvector operators $O_9$ and $O_{10}$, respectively.\cite{RMP68-1125-1996}
The Wilson coefficients are experimental observables.
Contributions from new physics appear in the experiment as deviations
from the SM values, which have been calculated to next-to-next-to-leading
order (NNLO).

Our experimental knowledge on the Wilson coefficients comes from 
the inclusive $b\rightarrow s\gamma$ branching fraction 
(Sec.~\ref{subsec:b2sgincl}), which determines 
the absolute value of $C_7$ to about 20\% accuracy, but not its sign, 
and from the inclusive $b\rightarrow s\ell^+\ell^-$ branching fraction,
which constrains $C_9$ and $C_{10}$ to an annular region in the
$C_9$-$C_{10}$ plane,\cite{PRD66-034002-2002} but gives no information
on the individual signs and magnitudes of these coefficients.
To further pin down the values of these coefficients, it is necessary
to exploit interference effects between the contributions from
different operators.
This is possible in $b\rightarrow s\ell^+\ell^-$ decays by evaluating
the differential inclusive decay rate as a function of the lepton invariant
mass, $m_{\ell\ell}^2 = q^2$ (Sec.~\ref{subsec:b2sllincl}), or by 
measuring the forward-backward asymmetry in the exclusive decay
$B\rightarrow K^*\ell^+\ell^-$ (Sec.~\ref{subsec:AFB}).
 
\subsection{$b\rightarrow s\ell^+\ell^-$ inclusive}\label{subsec:b2sllincl}

Measurements of the inclusive $b\rightarrow s\ell^+\ell^-$ decay rate have
been published by Belle~\cite{PRD72-092005-2005} and \babar,\cite{PRL93-081802-2004}
who also reports a direct CP asymmetry compatible with zero.
The partial $b\rightarrow s\ell^+\ell^-$ decay rate in the lepton invariant
mass range below the $J/\psi$ resonance is sensitive to the sign of 
$C_7$.\cite{PRD55-4273-1997}
A recent compilation~\cite{PRL94-061803-2005} of 
Belle~\cite{PRD72-092005-2005} and \babar~\cite{PRL93-081802-2004} data
shows that the currently available data clearly favour a negative sign 
for $C_7$, as predicted by the SM.

\subsection{$b\rightarrow s\ell^+\ell^-$ exclusive}\label{subsec:b2sllexcl}

Five of the eight individual $B\rightarrow K^{(*)}\ell^+\ell^-$ modes
have been established by now, the exceptions being $K^0e^+e^-$, 
$K^{*+}e^+e^-$ and $K^{*+}\mu^+\mu^-$. 
The charge and lepton-flavour averaged branching fractions obtained by 
\babar\ are~\cite{PRD73-092001-2006}
$\mathcal{B}(B\rightarrow K\ell^+\ell^-)   = 0.34\pm0.07\pm0.02$ and 
$\mathcal{B}(B\rightarrow K^*\ell^+\ell^-) = 0.78^{+0.19}_{-0.17}\pm0.11$,
the Belle results read~\cite{hep-ex/0410006}
$\mathcal{B}(B\rightarrow K\ell^+\ell^-)   = 0.550^{+0.075}_{-0.070}\pm0.027$ and 
$\mathcal{B}(B\rightarrow K^*\ell^+\ell^-) = 1.65^{+0.23}_{-0.22}\pm0.10$,
where we have combined systematic and model-dependence errors.
The striking phenomenon that the Belle values are higher by almost a factor of
two with respect to \babar's is present in all individual modes with significant
yields, but is probably attributable to statistics.
Both experiments have searched for asymmetries with respect to lepton flavour,
charge and isospin, so far without finding any surprises.

\subsection{Forward-backward asymmetry in $B\rightarrow K^*\ell^+\ell^-$}
  \label{subsec:AFB}

The forward-backward asymmetry in $B\rightarrow K^*\ell^+\ell^-$ decays is
defined as
\be
 A_\mathrm{FB}(q^2) = \frac{\Gamma(q^2, \cos\theta_{B\ell^-} > 0) -
                     \Gamma(q^2, \cos\theta_{B\ell^-} < 0)}
                    {\Gamma(q^2, \cos\theta_{B\ell^-} > 0) +
                     \Gamma(q^2, \cos\theta_{B\ell^-} < 0)},
\ee
where $\theta_{B\ell^-}$ is the angle between the momenta of the 
negative lepton and the $B$ meson in the dilepton rest frame
(positive lepton in the case of $\overline{B}$).
It is a non-trivial function of $q^2$ due to the interference between
vector ($C_7$, $C_9$) and axial-vector ($C_{10}$) couplings arising
from the relevant penguin and box diagrams.
In other words, by probing the interference between contributions from 
$\gamma$, $W$ and $Z$ exchanges, $A_\mathrm{FB}$ allows us to put to the test 
the very foundations of the Standard Model of electroweak interactions!

A recent analysis by Belle~\cite{hep-ex/0603018} using a 
data sample containing 386M 
$B\overline{B}$ pairs finds $114\pm13$ signal $B\rightarrow K^*\ell^+\ell^-$
decays with $K^*\rightarrow K^+\pi^-$, $K_S\pi^+$ and $K^+\pi^0$ and 
measures $A_\mathrm{FB}$ for the first time.
The integrated asymmetry is found to be $0.50\pm0.15\pm0.02$ ($3.4\sigma$ 
significance) and a fit to the double differential decay width 
$(1/\Gamma)d^2\Gamma/dq^2 d\cos\theta_{B\ell^-}$ on 8 event categories
(signal, 3 cross-feeds and 4 backgrounds) is used to extract ratios
of Wilson coefficients.
To facilitate comparison with various extensions of the SM, the evaluation 
is done for the leading-order terms $A_7$, $A_9$ and $A_{10}$ of the 
Wilson coefficients, thus assuming that the higher-order corrections
are the same as in the SM.
Since $A_7$ is known experimentally up to a sign from the 
$b\rightarrow s\gamma$ branching fraction and $A_\mathrm{FB}$ is not sensitive to 
that parameter, it is fixed to its (experimentally confirmed) SM 
value~\cite{PLB507-162-2001} in the fit ($A_7=-0.330$), and the results 
are expressed as the ratios
$A_9/A_7 = -15.3^{+3.4}_{-4.8}\pm1.1$ (SM: $-12.3$) and 
$A_{10}/A_7 = -10.3^{+5.2}_{-3.5}\pm1.8$ (SM: $12.8$),
in excellent agreement with the SM values given in brackets.
The same fit with a sign-flipped $A_7$ gives very similar 
results.
For $A_7$ left free within the experimentally allowed region,
the product of the two ratios is constrained to be in the interval
$-1.40\times 10^3 < A_9A_{10}/A^2_7 < -26.4$ at 95\% confidence level,
i.e.\ new physics scenarios with positive $A_9A_{10}$ are excluded
at such confidence.
As a cross-check Belle has also measured the forward-backward asymmetry
in the decay $B^+\rightarrow K^+\ell^+\ell^-$, for which no asymmetry 
is expected (no vector interference). 
The measured integrated asymmetry in that channel,
$0.10\pm0.14\pm0.01$, is indeed compatible with zero.

At this conference \babar\ has released its first measurements of 
angular distributions in $B\rightarrow K^{(*)}\ell^+\ell^-$ decays
using a data sample comprising 229M $B\overline{B}$ decays.
The study includes CP and lepton asymmetries, as well as a measurement
of the $K^*$ longitudinal polarization in $B\rightarrow K^*\ell^+\ell^-$ 
decays.
For the details and results of this analysis we refer to
the corresponding \babar\ publication,\cite{PRD73-092001-2006} 
which has become available shortly after the conference.

\subsection{$b\rightarrow s\nu\bar{\nu}$}\label{subsec:b2snunu}

The transition analog to $b\rightarrow s\ell^+\ell^-$ with neutral 
leptons in the final state, $b\rightarrow s\nu\bar{\nu}$, is theoretically
much cleaner than its charged counterpart, thanks to the absence of
the photon penguin diagram and hadronic long-distance effects (charmonium
resonances).
From the experimental point of view the modes mediated by 
$b\rightarrow s\nu\bar{\nu}$ are extremely challenging due to the presence
of two neutrinos in the final state.
Searches for such modes at the $B$ factories are therefore based on the 
so-called ``recoil method'': events are selected in which one $B$ meson is 
fully reconstructed in a hadronic or semileptonic mode. 
These events then provide an extremely clean environment to search for the
decay in question, at the expense of a rather low efficiency.

The only exclusive decay of this category that has been searched for
at the $B$ factories is $B^+\rightarrow K^+\nu\bar{\nu}$.
Using a sample of 89M $B\overline{B}$ pairs, \babar\ sets the limit
$\mathcal{B}(B^+\rightarrow K^+\nu\bar{\nu})<52$ on the branching
fraction.\cite{PRL94-101801-2005} 
The analysis uses about 
480k $B^+\rightarrow D^{(*)0}\ell^+\nu$ and  
180k $B^+\rightarrow D^{(*)0}X_\mathrm{had}^+$ decays, 
where $X_\mathrm{had}^+$ stands for up to five pions or kaons.
After finding exactly one opposite-charged kaon in the event remainder,
the primary selection criterion is a limit on the extra energy found
in the electromagnetic calorimeter, $E_\mathrm{extra} < 200$ MeV.
The Belle search~\cite{hep-ex/0507034} for $B^+\rightarrow K^+\nu\bar{\nu}$ 
is essentially a by-product of their search for $B^+\rightarrow \tau^+\nu_\tau$, 
which we will describe in Sec.~\ref{subsec:B2lnu}.
The limit obtained with 275M $B\overline{B}$ pairs is
$\mathcal{B}(B^+\rightarrow K^+\nu\bar{\nu})<36$.
The experimental limits are still an order of magnitude away from the
SM value,\cite{PRD63-014015-2001} 
$\mathcal{B}(B^+\rightarrow K^+\nu\bar{\nu}) = 3.8^{+1.2}_{-0.6}$.

\section{Annihilation tree and penguin decays to leptons and photons}
\label{sec:annpeng}


\subsection{$B^+\rightarrow \ell^+\nu$}\label{subsec:B2lnu}

The leptonic decays of charged $B$ mesons give direct access to
the product of the decay constant $f_B$ and the CKM-matrix element
$V_{ub}$, according to
\be \label{eq:B2lnu}
\mathcal{B}(B^+\rightarrow \ell^+\nu_\ell) =
\frac{G_F^2 m_B}{8\pi} m_\ell^2 \left(1-\frac{m_\ell^2}{m_B^2}\right)^2
f_B^2 |V_{ub}|^2 \tau_B .
\ee
Allowing for decay amplitudes beyond the SM, measurements of
these decays give stringent limits on important parameters of such
SM extensions, e.g.\ the mass of the charged Higgs boson and 
$\tan\beta$ in the minimal supersymmetric SM, or leptoquark masses in 
Pati-Salam models.

The most recent efforts have concentrated on $B^+\rightarrow \tau^+\nu_\tau$,
where an observation seems tantalizingly close.
The presence of one or more neutrinos in the final state again implies
that experimental searches are limited to the recoils of fully reconstructed
$B$ decays.
The Belle analysis~\cite{hep-ex/0507034} starts from a sample of 
about 400k $B^+\rightarrow D^{(*)0}h^+$ and $D^{(*)0} D_s^{(*)+}$ 
events ($h = \pi, K$) selected in 275M $B\overline{B}$ pairs,
whereas  \babar~\cite{PRD73-057101-2006} uses a similar number of semileptonic
$B^+\rightarrow D^{(*)0}\ell^+\nu$ events (from 232M $B\overline{B}$ pairs).
Decays of the types $\tau\rightarrow \mu(e)\nu\bar{\nu}$, $\pi\nu$, $\pi\pi^0\nu$, 
and $\pi\pi\pi\nu$ are then searched for in the event remainders.
The final event selection is based on the extra energy present in the
electromagnetic calorimeter. 
So-called double-tag events (i.e.\ fully reconstructed $\Upsilon(5S)$ 
decays) are used to validate the simulation of this quantity.
The limits obtained are 180 
(Belle~\footnote{A few weeks after this conference, the Belle collaboration has
announced evidence for $B^+\rightarrow \tau^+\nu_\tau$.\cite{hep-ex/0604018}}) 
and 260 (\babar, combined with a previous
analysis~\cite{PRL95-041804-2005} based on hadronic $B$ decays).
Both experiments report positive, but still insignificant mean values,
81$^{+58}_{-45}$ and 130$^{+58}_{-45}$, respectively, which HFAG averages
to 92$^{+51}_{-41}$, a number which coincidentally lies very close
to the SM prediction~\cite{PRD73-057101-2006} of $93\pm39$. 

The related decays $B^+\rightarrow \mu^+\nu_\mu$ and $B^+\rightarrow e^+\nu_e$
are helicity suppressed with respect to $B^+\rightarrow \tau^+\nu_\tau$ by
factors of 223 and 10$^7$, respectively (Eq.~\ref{eq:B2lnu}).
No new limits on these decay channels have been reported since 2004.
For the sake of completeness, we note that the best limits so far have
been reported (but not published) by Belle, 
$\mathcal{B}(B^+\rightarrow \mu^+\nu_\mu) < 2.0$ 
(152M $B\overline{B}$)~\cite{hep-ex/0408132} and
$\mathcal{B}(B^+\rightarrow e^+\nu_e) < 5.4$
(65M $B\overline{B}$).\cite{BELLE-CONF-0247}
Belle also gives~\cite{hep-ex/0408132}
$\mathcal{B}(B^+\rightarrow \mu^+\nu_\mu\gamma) < 23$ and 
$\mathcal{B}(B^+\rightarrow e^+\nu_e\gamma) < 22$.
\babar\ has published the limit~\cite{PRL92-221803-2004}
$\mathcal{B}(B^+\rightarrow \mu^+\nu_\mu) < 6.6$. 

\subsection{$B^0\rightarrow \ell^+\ell^-$}\label{subsec:B2ll}

An important advance in this category has been achieved by \babar\ by
establishing the very first limit~\cite{hep-ex/0511015} on the decay 
$B^0\rightarrow \tau^+\tau^-$.
This previously unconstrained decay represented a big loophole for 
theorists.\cite{PRD55-2768-1997} 
An experimental limit constrains in particular leptoquark couplings
and $\tan\beta$ enhancements in Supersymmetry.
The 2--4 neutrinos in the final state render the experimental
search extremely difficult.
The \babar\ analysis starts from 280k fully reconstructed $B^0\rightarrow
D^{(*)}X$ decays, where $X$ stands for a combination of charged and neutral
pions and kaons.
Decay products of two ``simple'' $\tau$ decays are then searched for
in the event remainder, i.e.\ two $\tau$ decays with only one charged
particle each: $\tau\rightarrow \mu(e)\nu\bar{\nu}$, $\pi\nu$ and
$\rho\nu$, which together cover 51\% of all $\tau^+\tau^-$ decays.
After rejecting events containing identified neutral and charged kaons,
the kinematics of the charged daughters and
the residual energy in the electromagnetic calorimeter are fed into a 
neural network to separate signal from background.
The limit obtained in this way is 
$\mathcal{B}(B^0\rightarrow \tau^+\tau^-) < 3400$, 
four orders of magnitude above the SM value.

Concerning decays to pairs of lighter leptons, the best limits from the
$B$ factories come from \babar, 
$\mathcal{B}(B^0\rightarrow \mu^+\mu^-) < 0.083$ and  
$\mathcal{B}(B^0\rightarrow e^+e^-) < 0.061$, obtained with
120M $B\overline{B}$ pairs,\cite{PRL94-221803-2005} representing
an improvement of over a factor of two with respect to the previous
Belle limits.\cite{PRD68-111101-2003}
The Tevatron experiments have now taken the lead in the search for
$B\rightarrow \mu^+\mu^-$:
CDF reports the limits~\cite{PRL95-221805-2005} 
$\mathcal{B}(B^0\rightarrow \mu^+\mu^-) < 0.039$ and  
$\mathcal{B}(B^0_s\rightarrow \mu^+\mu^-) < 0.15$ obtained with
364 pb$^{-1}$ of data, while
D$\emptyset$ (not having sufficient mass resolution to distinguish
$B^0$ from $B^0_s$) gives~\cite{D0-CONF-4733} 
$\mathcal{B}(B^0_s\rightarrow \mu^+\mu^-) < 0.3$ (300 pb$^{-1}$).
A combination of Tevatron data,\cite{hep-ex/0508058} taking into 
account common systematics, finds the limits
$\mathcal{B}(B^0\rightarrow \mu^+\mu^-) < 0.032$ and  
$\mathcal{B}(B^0_s\rightarrow \mu^+\mu^-) < 0.12$, which
already put stringent constraints~\cite{JHEP0509-029-2005} 
on some supersymmetric models!

\subsection{$B^0\rightarrow \nu\bar{\nu}$}\label{subsec:B2nn}

The decay $B^0\rightarrow \nu\bar{\nu}$ is essentially forbidden in the SM,
hence any evidence for invisible decays of $B$ mesons would point to 
exotic phenomena such as neutralinos or large extra dimensions.
The only limit so far is the one published by \babar,\cite{PRL93-091802-2004}
$\mathcal{B}(B^0\rightarrow \nu\bar{\nu}) < 220$. 
It is based on the analysis of the recoils of 126k reconstructed 
$B\rightarrow D^{(*)}\ell\nu$ events (from 88.5M $B\overline{B}$ pairs).
By looking for one energetic photon in the same events, \babar\ also obtains
a limit for the associated radiative decay,
$\mathcal{B}(B^0\rightarrow \nu\bar{\nu}\gamma) < 47$. 

\subsection{$B^0\rightarrow \gamma\gamma$}\label{subsec:B2gg}

Like the other decays mediated by annihilation diagrams, 
the purely radiative decay of the $B^0$ could receive enhancements from
the exchange of charged Higgs bosons or more exotic charged particles.
Belle has recently published its first search for this channel, based on 
a data sample equivalent to 111M $B\overline{B}$ pairs.\cite{PRD73-051107-2006}
The result,
$\mathcal{B}(B^0\rightarrow \gamma\gamma) < 0.62$, improves on a very
early \babar\ limit,\cite{PRL87-241803-2001} but is still about a 
factor 20 away from the SM prediction.

Applying a similar analysis to the 1.86 fb$^{-1}$ of data obtained from a 
short (3 days) engineering run at the $\Upsilon(5S)$, Belle has set a 
preliminary limit on the corresponding $B^0_s$ decay,\cite{Drutskoy} 
$\mathcal{B}(B^0_s\rightarrow \gamma\gamma) < 56$.
Note that the branching fraction predicted by the SM for this channel is 
around 1.2, a level that would be within reach of a few-months long run 
at the $\Upsilon(5S)$!

\section{Summary}
\label{sec:summary}

To summarize we note that in the past couple of years the availability of 
data samples containing several hundred million $B\overline{B}$ pairs at the $B$ 
factories has brought about decisive advances in the field of radiative 
and leptonic rare $B$ decays:
while the study of the $b\rightarrow s\gamma$ transition has turned into
a precision science, $b\rightarrow d\gamma$ has finally become observable;
in the $b\rightarrow s\ell^+\ell^-$ sector we have moved
from mere observation to the exploration of angular distributions probing
for the first time the Wilson coefficients at play, and 
we are at the brink of observing the first purely leptonic decay of
the $B$ meson, $B^+\rightarrow \tau^+\nu_\tau$.
On top of the statistics, the experience and expertise accumulated 
at the $B$ factories by now allow the experiments to tackle even the most 
challenging decay modes, as demonstrated by \babar's recent limit on 
$B^0\rightarrow \tau^+\tau^-$.

The wealth of new data not only curbs an array of 
new-physics models, but also begins to add significant constraints on the
CKM parameters
($|V_{ub}|$ from $B^+\rightarrow \tau^+\nu_\tau$, 
 $|V_{td}/V_{ts}|$ from $B\rightarrow [\rho,\omega]\gamma$ and
$B\rightarrow K^*\gamma$),
in addition and complementary to the ones obtained from hadronic 
decays.\cite{CKM}
So far, the SM still saves its bacon, but it may just be too early to
tell in these channels, which remain our most promising scouts for
new physics beneath the energy frontier.


\section*{Acknowledgments}
{
I am indebted to all the members of the \babar\ and Belle collaborations who 
have contributed to this review in one way or another.
In particular I thank R.~Faccini and J.~Berryhill (\babar), as well as
M.~Nakao and K.~Ikado (Belle) for their invaluable help.   
This work was supported by the Swiss National Science Foundation under grant
Nr.~620-066162.
}

\end{document}